# MACHINE-LEARNING TECHNIQUES FOR THE OPTIMAL DESIGN OF ACOUSTIC METAMATERIALS

A. Bacigalupo[1], G. Gnecco[1], M. Lepidi[2], and L. Gambarotta[2]

[1] MUSAM and AXES Research Units - IMT School for Advanced Studies
Piazza S. Francesco, 19 - 55100 Lucca, Italy
e-mail: {andrea.bacigalupo, giorgio.gnecco}@imtlucca.it

[2] Department of Civil, Chemical, and Environmental Engineering - University of Genoa
Via Montallegro, 1 - 16145 Genoa, Italy
e-mail: {marco.lepidi, luigi.gambarotta}@unige.it

**Abstract.** Recently, an increasing research effort has been dedicated to analyse the transmission and dispersion properties of periodic acoustic metamaterials, characterized by the presence of local resonators. Within this context, particular attention has been paid to the optimization of the amplitudes and center frequencies of selected stop and pass bands inside the Floquet-Bloch spectra of the acoustic metamaterials featured by a chiral or antichiral microstructure. Novel functional applications of such research are expected in the optimal parametric design of smart tunable mechanical filters and directional waveguides. The present paper deals with the maximization of the amplitude of low-frequency band gaps, by proposing suitable numerical techniques to solve the associated optimization problems. Specifically, the feasibility and effectiveness of Radial Basis Function networks and Quasi-Monte Carlo methods for the interpolation of the objective functions of such optimization problems are discussed, and their numerical application to a specific acoustic metamaterial with tetrachiral microstructure is presented. The discussion is motivated theoretically by the high computational effort often needed for an exact evaluation of the objective functions arising in band gap optimization problems, when iterative algorithms are used for their approximate solution. By replacing such functions with suitable surrogate objective functions constructed applying machine-learning techniques, well-performing suboptimal solutions can be obtained with a smaller computational effort. Numerical results demonstrate the effective potential of the proposed approach. Current directions of research involving the use of additional machine-learning techniques are also presented.

# 1. Introduction

The extraordinary perspectives offered by the recent theoretical advances and technological developments in the field of micro-architected materials have attracted a growing attention from a large and multi-disciplinary scientific community [1],[2],[3]. In particular, an increasing research effort has been dedicated over the last years to analyze the transmission and dispersion properties of elastic waves propagating in periodic microstructures characterized by the presence of one or more auxiliary oscillators per elementary cell [4],[5],[6]. Such periodic systems – known as *acoustic* metamaterials – can leverage the local resonance mechanism between the cellular microstructure and the auxiliary oscillators (*resonators*) to ensure a variety of functional improvements with respect to the corresponding (resonator-free) materials. Indeed, if properly designed according to a suited *parametric optimization* strategy (see [7]), the optimized geometric, inertial and elastic parameters of the periodic microstructure and of the resonators may allow the achievement of desirable wave dispersion properties. Consequently, challenging issues in the modern field of theoretical and applied optimization arise in optimally designing the spectral metamaterial properties for specific objectives, including – for instance – opening, enlarging, closing or shifting band gaps in desired acoustic frequency ranges. From the technological viewpoint, a successful optimization of the dispersion properties for acoustic metamaterials paves the way for developing a new generation of smart engineering devices, such as directional waveguides, mechanical filters, negative refractors, sub-wavelength edge detectors, invisibility cloaks, ultrasound focusers, and mechanical energy propagators [8],[9],[10],[11],[12].

A promising technique for the spectral design of acoustic metamaterials is based on the formulation and numerical solution of constrained nonlinear optimization problems, typically characterized by continuous optimization variables. Such optimization problems can be suitably stated as follows. First, some geometric, elastic and inertial properties (such as the mass and stiffness of the local resonators) are assumed as design parameters, playing the role of optimization variables. Second, mechanical and geometric limitations (such as validity boundaries for the assumptions of the model) are taken into account to establish the linear/nonlinear constraints. Third, any desirable physical property associated with the metamaterial spectrum is adopted as nonlinear objective function. Depending on functional requirements, the objective function can be defined by properly weighting different spectral properties [13],[14],[15],[16],[17], or by combining dynamic and static features [18]. Unfortunately, the numerical solution of such optimization problems by straightforwardly applying classical iterative algorithms [19] tends to become highly demanding and, consequently, often inconvenient. Specifically, a first time-consuming bottleneck in the application of such algorithms can be identified in the evaluation of the objective function at each iteration. Indeed, evaluating the objective function typically requires the numerical solution of a sequence of eigenvalue or generalized eigenvalue subproblems, one for each value of the (discretized) domain of the dispersion functions. A second, similar issue may rise up for the approximate computation of the partial derivatives of the objective functions with respect to the optimization variables, whenever the iterative algorithm exploits such information during its iterations. Finally, it is worth remarking that the computational effort rapidly grows up with (i) the increment of the physical model dimensions (larger eigenspace and, usually, higher spectral density) and/or (ii) the increase of the number of design parameters taken as optimization variables (larger domain of the objective function).

The computational hurdles slowing down the spectral optimization process can be by-passed by suitably reformulating the problem in a two-phase procedure. By replacing the original (or *true*) objective functions (and their partial derivatives) with more-easily computable approximations, surrogate optimization [20],[21] can tackle the spectral optimization problems in a numerically efficient way, rapidly providing well-performing suboptimal solutions (first phase). Therefore, if necessary, the suboptimal solutions could be re-optimized locally, turning back to the true objective functions, hence to the original optimization problems (second phase).

Within this challenging framework, the present contribution is focused on the formulation and application of surrogate optimization techniques - based on interpolation through *Gaussian Radial Basis Function* (*RBF*) *networks* and *Quasi-Monte Carlo sequences* - for the optimization of the filtering properties of a periodic acoustic metamaterial with tetrachiral cellular topology. Based on the physical-mathematical lagrangian model governing the free propagation of elastic waves through the metamaterial [22], the paper intends to thoroughly revise the methodological approach presented in [23], by conveniently re-formulating and extending it in the framework of a surrogate optimization strategy. To this purpose, proper motivations for the application of interpolation techniques in the construction of surrogate objective functions to be used in band gap optimization problems are provided first (Section 2). Second, the new approach adopted for the numerical solution of these optimization problems is detailed (Section 3), with focus on the combination of Gaussian RBF networks, a suitable iterative optimization algorithm, and Quasi-Monte Carlo sequences. The latter, in particular, are applied for both the construction of the Gaussian RBF networks and the re-initialization of the optimization algorithm. Then, the numerically optimized designs obtainable for the microstructure of the acoustic tetrachiral metamaterial are presented, and properly discussed with reference to the existent literature (Section 4). Original ideas and viable directions for further research developments, based on the use of additional machine-learning techniques are finally outlined (Section 5). In order to favour the paper readability for the largest possible audience, some specific mathematical concepts, not strictly necessary to present and discuss the principal substantial aspects of the research, have been organized in three thematic Appendices, in which formal notations and complete technical details are reported.

## 2. Motivations for machine learning in spectral design problems

A fundamental problem in the spectral design of periodic materials and metamaterials consists in the maximization of the stop bandwidth (amplitude of the band gap) separating a selected pairs of consecutive dispersion curves in the acoustic part of the dispersion spectrum [24],[25],[26],[27]. From the mathematical viewpoint, the matter can be approached by formulating a constrained nonlinear optimization problem in the form

$$\begin{aligned}&\underset{\mathbf{x}\in\mathbb{R}^p}{\text{maximize}} f(\mathbf{x}),\\ &\text{s.t.} \quad h_i(\mathbf{x})=0,\ i=1,...,n_{eq},\\ &\qquad g_j(\mathbf{x})\leq 0,\ j=1,...,n_{ineq},\end{aligned} \quad (1)$$

where $\mathbf{x} \in \mathbb{R}^p$ is the vector collecting the $p$ real-valued optimization variables and $f(\mathbf{x})$ is the objective function. The auxiliary functions $h_i(\mathbf{x})$ and $g_j(\mathbf{x})$ are employed to express the constraints in the mathematical form of equalities and inequalities, respectively.

Considering a lagrangian beam lattice formulation, a $M$-dimensional physical-mathematical model can be employed to govern the free propagation of linear elastic waves in the periodic tetrachiral metamaterial [22]. Denoting by $\boldsymbol{\mu}$ the vector collecting the minimal set of independent dimensionless mechanical parameters of the lagrangian model, and denoting by $\mathbf{k}$ the dimensionless wavevector spanning the Brillouin zone of the periodic lattice, the Floquet-Bloch spectrum is characterized by $M$ real-valued dispersion relations $\omega(\boldsymbol{\mu},\mathbf{k})$. Such relations satisfy the characteristic equation $F(\omega,\boldsymbol{\mu},\mathbf{k})=0$ associated with the linear eigenproblem governing the free-wave propagation according to the Floquet-Bloch theory. Therefore, the objective function $f(\mathbf{x})$ describing the amplitude of a certain spectral band gap in the spectral design optimization problem (1) can be established as a suited combination of two consecutive dispersion relations $\omega(\boldsymbol{\mu},\mathbf{k})$. Consequently, all (or part of) the mechanical parameters in the $\boldsymbol{\mu}$-vector play the role of optimization variables, or design parameters, composing the vector $\mathbf{x}$.

The main computational issues arising when trying to solve numerically this kind of optimization problems can be discussed. The evaluation of the band gap amplitude associated with any given choice for the set of design parameters requires two main computational steps:

a) determining the $M$ coefficients of the characteristic polynomial $F(\omega,\boldsymbol{\mu},\mathbf{k})$ related to the determinant of a square ($M$-by-$M$) Hermitian matrix $\mathbf{H}(\omega,\boldsymbol{\mu},\mathbf{k})$, and then calculating the $M$ polynomial roots. This step has to be repeated several times, each time considering a different wavevector $\mathbf{k}$ identifying one of the points necessary to discretize with sufficient fineness the closed boundary $\partial B$ of a suitable subdomain $B$ of the Brillouin zone;

b) determining the maximum and minimum values of two selected consecutive dispersion curves, and then comparing these values.

Within this two-step procedure, the computational effort is clearly dominated by the first step, and increases by increasing either the dimension of the square matrix (model improvement), or the refinement of the discretization of the closed boundary $\partial B$. It is worth remarking that, in this context, a symbolic computation of the coefficients of the characteristic polynomial does not help to decrease the computational effort (actually, it may even increase it), unless the square matrix $\mathbf{H}(\omega,\boldsymbol{\mu},\mathbf{k})$ is sparse, which is usually not the case in this kind of physical models. Only for a sparse matrix, indeed, a significant reduction in the number of terms appearing in the symbolic expression of the matrix determinant is expected. In all the other cases, a numerical evaluation of the coefficients for each choice of the $\boldsymbol{\mu}$-vector of design parameters has to be preferred. However, a more feasible way to reduce – possibly significantly – the computational effort consists in reducing the number of evaluations of the objective function of the original optimization problem. This computational cost-saving target can be achieved by introducing acceptable simplifications, including – for instance – the adoption of machine-learning techniques for function interpolation or approximation. It has to be remarked that the considerations motivating the employment of machine-learning techniques are quite general, and their validity holds also for other optimizable topologies of periodic metamaterials [15],[16],[17].

## 3. Surrogate optimization in spectral design

In order to decrease the computational effort associated with the numerical solution of band gap optimization problems, a *surrogate optimization* approach can be considered. The leading idea is that the true objective function can be evaluated exactly on a finite subset of the domain to which the set of continuous design variables belongs. In machine-learning terms, this subset is called *training set*. Then, the function values are interpolated/extrapolated in the remaining domain of the design variables. As a consequence, the numerical optimization is performed by replacing the true objective function with its interpolant, whose evaluation is expected to be easier (once the interpolant has been constructed in the training phase). According to this general strategy, the approach proposed in this paper to tackle the band gap optimization problems under investigation is characterized by the following features:

a) a *mesh-free interpolation* is used for interpolating the objective function values, using a finite number $N_c$ of *strictly positive-definite* RBF computational units [28]. Technical details about the mesh-free interpolation are reported in Appendix A. Moreover, a *Quasi-Monte Carlo discretization* [29] of the domain of design variables is exploited to select the centers of the RBFs, after removing from the Quasi-Monte Carlo sequence all the points that do not satisfy the constraints of the original optimization problem. These choices are motivated by the nice approximation error guarantees (expressed in the sup norm), that are available for the mesh-free interpolation method, when both the objective function to be interpolated is smooth enough and the fill distance is small enough (see, for instance, Theorem 14.5 in [28]). Loosely speaking, the fill distance is a way to quantify how well the training points fill the optimization domain. A small value of the fill distance can be obtained, e.g., when the training points are generated using a (mesh-free)

Quasi-Monte Carlo discretization [29]. In the paper, *Gaussian RBFs* with identical width are used for the interpolation. The width is chosen by *leave-one-out cross-validation*, following the approach described in [28] (in Section 17.1.3) and therein called *Rippa method* [30]. For simplicity, the centers of the Gaussian RBF computational units are chosen coincident with the training points. In this way, the strict positive-definiteness of the Gaussian RBF computational units guarantees the non-singularity of the associated *interpolation matrix*, hence, also the uniqueness of the interpolant (see Chapter 3 in [28]);

b) a *surrogate objective function* is constructed by using the *mesh-free function interpolation method*, because the values assumed by the true objective function at the input training points are known exactly. This specific point distinguishes the approach from alternative machine-learning methods like *Support Vector Regression* (see for instance [31] in Section 6.2), which exploit, instead, function approximation. Technical details about the mesh-free interpolation are reported in Appendix B;

c) a numerical optimization of the surrogate objective function is performed by applying a *Sequential Linear Programming* (*SLP*) iterative algorithm based on an *adaptive trust region*. This version of the SLP algorithm has some similarities with that presented in [32] for a different application of numerical optimization in materials science. At each iteration of the SLP algorithm, the initial optimization problem (whose objective function is the surrogate one) is replaced by its linearization around the current vector of design variables. Changes in these variable values are also limited by the presence of an additional adaptive trust region constraint, which typically depends on the quality of the linearization performed in the previous iterations. Specifically, at each iteration but the first one, the size of the trust region (which is centered on the current choice for the vector of design variables) is expanded (up to an upper bound on that size) when the previous linear approximation is accurate. This means that all the relative errors between the linear approximations of the partial derivatives of the objective function (and possibly, also of the constraining functions, when they are nonlinear) and their true values at the current point are smaller than a given tolerance $\varsigma > 0$. Otherwise, the size of the trust region is reduced (up to a lower bound). A possible choice for the adaptive trust region is a hypercube with edges of length $L$, whose lower and upper bounds are denoted, respectively, by $L_{\min}$ and $L_{\max}$. Its initial length is denoted by $L_{\text{init}}$. Here, the SLP algorithm is terminated after a fixed number $N_i$ of iterations, although more sophisticated termination criteria may be also exploited. Differently from [15],[16],[17],[22], the SLP algorithm is used instead of the *Globally Convergent* version of the *Method of Moving Asymptotes* (*GCMMA*) [33],[34], since some preliminary simulations suggest that, at least for the specific optimization problem, the performance of SLP does not depend strongly on the interpolation quality. Technical details about the mesh-free interpolation are reported in Appendix C;

d) an iterative optimization algorithm is applied for a finite number $N_s$ of times, by adopting a *Quasi-Monte Carlo multi-start initialization approach* (identical to that employed in [15],[16],[17],[22]), which produces $N_s$ different starting points. In addition, a subset of the same input training points used for the construction of the interpolant is selected also to perform the Quasi-Monte Carlo multi-start initialization, since on such points the surrogate objective function reproduces exactly the true objective function, committing a zero approximation error therein. Basically, at each initial iteration of the algorithm, the true objective function and the surrogate one assume the same value.

It is worth remarking that the proposed approach has some similarities with the methodology applied in [35] to a mechanical design problem. The similarities include the application of an RBF network interpolant in the context of surrogate optimization, whereas the differences consist of (i) the use of a Quasi-Monte Carlo sequence both for constructing the interpolant and for initializing the optimization, (ii) the application of leave-one-out cross-validation to select the common width parameter of the Gaussian RBFs, and (iii) the use of the SLP algorithm with the adaptive trust region for the numerical optimization.

As final remark, the maximum value achieved by the Gaussian RBF network interpolant on the whole domain of design variables is typically different from that assumed by the surrogate/true objective function in correspondence of the training set. This occurrence can be justified on the light of the additional simplifying condition that all the inequality constraints are inactive at optimality (i.e., that they hold with the strict inequality). When this condition holds, a necessary (but not sufficient) condition for getting the same maximum values in the two cases is that the gradient of the Gaussian RBF network interpolant is equal to the all-zero vector, for the elements of the training set associated with the maximum objective value on such a set. However, this condition, combined with the interpolating conditions, establishes a system of linear equations having typically no solution (instead, the interpolating conditions alone always determine a unique Gaussian RBF network interpolant). Consequently, the maximum value of the Gaussian RBF network interpolant on the whole domain of design variables is typically larger than the maximum value of the true objective function on the set of input training points.

**4. Spectral design of the tetrachiral material**

The surrogate optimization method can be applied to the design problem of maximizing the band gap amplitude in the low-frequency range of the dispersion spectrum associated with the tetrachiral periodic metamaterial. Following the operational procedure outlined in Section 3, the Gaussian RBF network interpolant of the true objective function is constructed using a training set made of $N_c = 500$ points, generated starting from a *Sobol' sequence* [29],[36]. The points violating the constraints of the optimization problem are discarded. The common width parameter of the Gaussians is tuned automatically using leave-one-out cross-validation. Then, the numerical optimization is performed by executing $N_s = 10$ times the SLP algorithm with the adaptive trust region, initializing that algorithm each time with a different center. Each repetition consists of $N_i = 100$ iterations of the SLP algorithm. The adaptive trust region is chosen as a hypercube with edges of length $L$, centered on the parameter vector generated at each iteration. The initial length is $L_{init} = 0.1$, whereas its lower and upper bounds are, respectively, $L_{min} = 0.005$ and $L_{max} = 1$. At the end of each iteration, the value of the current length $L$ is doubled (respectively, reduced by one half), up to its upper (respectively, lower) bound, depending on the linearization accuracy, which is measured by the tolerance $\varsigma = 0.1$.

Considering the lagrangian physical-mathematical model of the tetrachiral metamaterial [12],[22], a set of four nondimensional mechanical parameters are selected as design variables, and collected in the vector $\boldsymbol{\mu} = (w_{an}/w, E_r/E_s, v_r, \rho_r/\rho_{an}) \in \mathbb{R}^4$. With reference to the characteristic microstructure of the tetrachiral periodic cell (Figure 5a), the geometric quantities $w$ and $w_{an}$ stand for the transversal widths of the beams and rings, respectively. The elastic properties $E_s, E_r$ and $v_r$ are the Young moduli and the Poisson ratios of the beams and rings (subscript $s$) and the resonator (subscript $r$). Finally, the inertial properties $\rho_r$ and $\rho_{an}$ represent the different mass densities of the resonator and ring, respectively. The four design variables have been selected as the most convenient optimizable quantities among the complete set of independent mechanical parameters, according to the outcomes of previous, effective but less refined, optimization strategies based on an eight-parameter set [22]. For the sake of completeness, the remaining four (fixed) parameters are $w/\varepsilon = 3/50, R/\varepsilon = 1/5$, $r/\varepsilon = 3/50$ and $\beta = \arcsin(2/5)$, where $\varepsilon$ is the periodic cell length, $R$ and $r$ are the radii of the ring and the resonator, respectively, and finally $\beta$ is the chirality angle. It is worth remarking that the dimensional reduction in the design variable space simplifies the problem, without compromising the generality of the surrogate optimization method.

According to the selection of the design variables, the spectral optimization problem for the tetrachiral metamaterial can be focused on maximizing the amplitude of the band gap separating two low-frequency dispersion curves. Mathematically, the problem can be formulated as

$$\begin{aligned} \underset{\boldsymbol{\mu} \in \mathbb{R}^4}{\text{maximize}} \quad & \Delta \omega_{\partial B_1}(\boldsymbol{\mu}) \\ \text{s.t.} \quad & \mu_{l,\min} \leq \mu_l \leq \mu_{l,\max}, \quad l = 1,...,4, \end{aligned} \qquad (2)$$

where the $\boldsymbol{\mu}$-dependent objective function $\Delta \omega_{\partial B_1}(\boldsymbol{\mu})$ is the amplitude of the partial band gap between the second and third dispersion curves, properly normalized by dividing the amplitude by the center frequency of the band gap. The amplitude is evaluated as the difference between the maximum of the second frequency and the minimum of the third frequency over the horizontal segment $\partial B_1$ of the boundary $\partial B$ closing a suitable subdomain $B$ of the Brillouin zone. The amplitude can be positive or null, if the difference is negative (no band gap). Differently from [22], the problem constraints are limited to lower and upper bounds $\mu_{l.\min}$ and $\mu_{l.\max}$ imposed on the admissible range of each component $\mu_l$ of the $\boldsymbol{\mu}$-vector (see Table 1).

Table 1. Lower and upper bounds on the geometrical and mechanical parameters.

|  | $w_{an}/w$ | $E_r/E_s$ | $v_r$ | $\rho_r/\rho_{an}$ |
|---|---|---|---|---|
| $\mu_{l.\min}$ | 1/3 | 1/10 | 1/5 | 1/2 |
| $\mu_{l.\max}$ | 10/3 | 1 | 2/5 | 2 |

Since two or more local maxima of the band gap amplitude can co-exist in the admissible range of the parameters, the largest amplitude among those found by the search algorithms (*best solution*) is assumed to actually solve the optimization problem. In the optimization problem (2), the band gap amplitude represents the true objective function $f(\boldsymbol{\mu})$, which in the framework of the surrogate optimization approach is instead replaced by its interpolant $f_{surr}(\boldsymbol{\mu})$. The best solution $\boldsymbol{\mu}^*_{surr}$ solving the surrogate optimization problem is compared with the best choice $\boldsymbol{\mu}^*_{train}$ of the parameter vector within the training set, i.e., that maximizing the surrogate objective function on that set (or, equivalently, that maximizing the true objective function therein, since the two functions coincide on the training set). For a further comparison, the best solution $\boldsymbol{\mu}^*_{true}$ achievable by means of the GCMMA optimization algorithm, applied to the true objective function using the same number of repetitions and starting from the same initializations as SLP applied to surrogate optimization, is also considered. Table 2 summarizes the best solutions, and the corresponding values of the surrogate/true objective function.

Table 2. Best solutions of the optimization problems and corresponding objective values.

|  | $w_{an}/w$ | $E_r/E_s$ | $v_r$ | $\rho_r/\rho_{an}$ | $f_{surr}(\boldsymbol{\mu})$ | $f(\boldsymbol{\mu})$ |
|---|---|---|---|---|---|---|
| $\boldsymbol{\mu}^*_{surr}$ | 0.333 | 0.100 | 0.400 | 2.000 | 0.482 | 0.450 |
| $\boldsymbol{\mu}^*_{train}$ | 0.410 | 0.166 | 0.212 | 1.735 | 0.295 | 0.295 |
| $\boldsymbol{\mu}^*_{true}$ | 0.333 | 0.100 | 0.280 | 2.000 | - | 0.447 |

The comparison of the solutions shows that the largest surrogate objective value $f_{surr}(\boldsymbol{\mu}^*_{surr})$ found by the surrogate optimization method is larger than the largest surrogate/true objective value $f_{surr}(\boldsymbol{\mu}^*_{train})$ achieved on the training set, demonstrating the usefulness of surrogate optimization for the specific application. This finding is aligned with the remarks in Section 3. Moreover, even though the surrogate value $f_{surr}(\boldsymbol{\mu}^*_{surr})$ actually overestimates the true value $f(\boldsymbol{\mu}^*_{surr})$, close objective values $f(\boldsymbol{\mu}^*_{surr}) \simeq f(\boldsymbol{\mu}^*_{true})$ are obtained. This results confirms that the surrogate optimization method can achieve nearly the same maximum objective value obtainable by the GCMMA-based maximization of the true objective function. In the particular case shown in Table 2, it provides even a slightly larger value of the true objective.

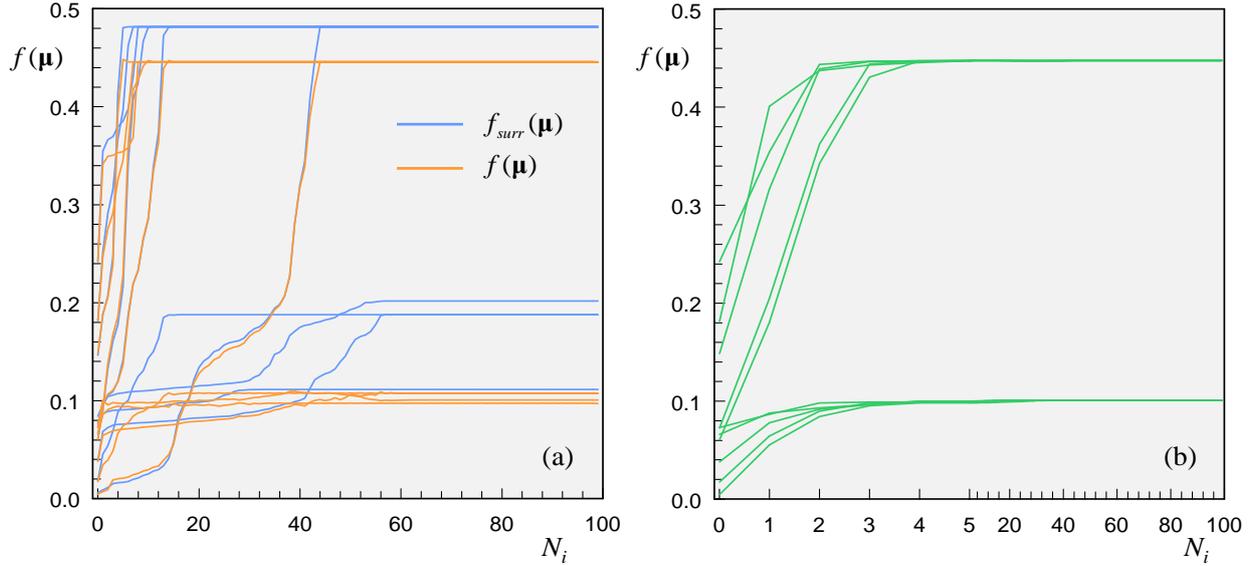

Figure 1. Objective function values versus iteration number: (a) comparison of the surrogate and true function values for the SLP algorithm applied to the surrogate objective function, (b) true function values for the GCMMA algorithm applied to the true objective function.

In order to deepen the analysis of the numerical results, Figure 1a illustrates, for each of the $N_s$ repetitions of the surrogate optimization (one line for each repetition), the values assumed by the surrogate objective function (blue lines) and by the true objective function (red lines). Both function values vary with respect to the iteration number, when a total number of $N_i = 100$ iterations of the SLP algorithm with the adaptive trust region is performed to solve numerically the surrogate optimization problem. It is worth remarking that these $N_i$ true objective values are not exploited by the SLP algorithm for further optimization, but are computed a-posteriori, for a final accuracy check. The numerical results reported in the figure fit the theoretical expectations, since the two objective functions evolve in a qualitatively similar way over the repetitions. This qualitative fitting persists even in scenarios of worse approximation by the Gaussian RBF network interpolant, possibly due to a sufficient but inadequate number of centers/interpolation points. For comparison, Figure 1b shows the results obtained by the GCMMA algorithm applied directly to the true objective function. It is worth noting that, at each iteration, GCMMA needs to compute both the true objective function and its gradient (i.e., four partial derivatives, which are approximated by finite differences). Hence, while the surrogate optimization method requires $N_c = 500$ evaluations of the true objective function (required to construct the interpolant), the GCMMA algorithm applied to the true objective function requires a one-order of magnitude larger number $N_r \cdot N_i \cdot (1+4) = 10 \cdot 100 \cdot 5 = 5000$ of evaluations. Further advantages of the surrogate optimization method are expected for higher-dimensional problems, for which the extra computation cost needed to construct the surrogate objective function can be compensated by a large cost of an exact evaluation of the true objective function [37].

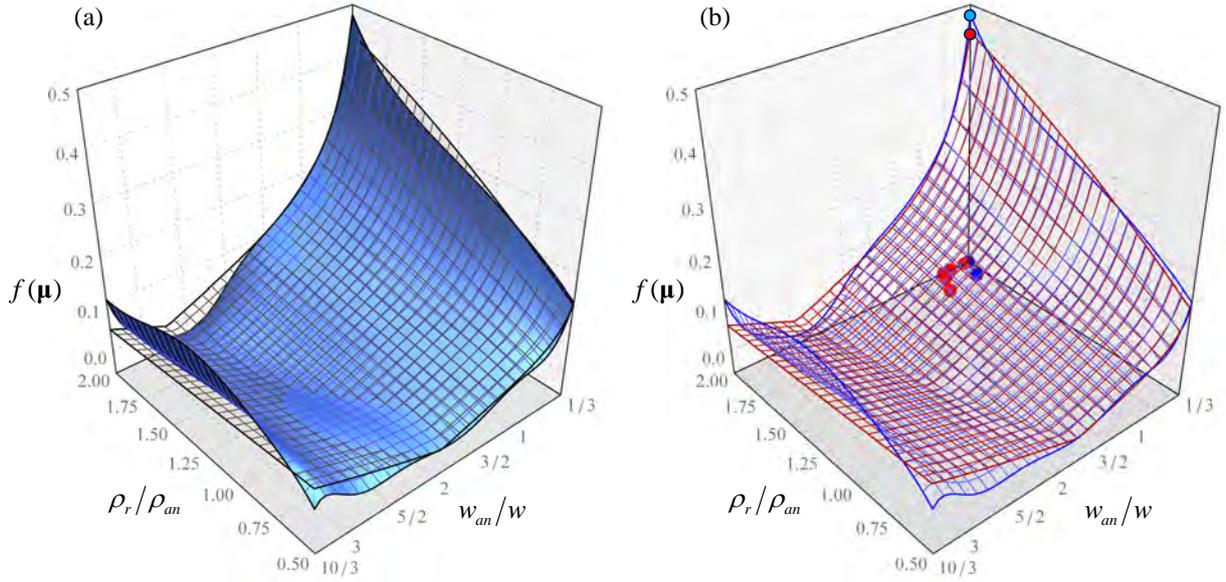

**Figure 2**. Objective functions in the bi-dimensional subspace of the admissible parameters $\rho_r/\rho_{an}$ and $w_{an}/w$: (a) surrogate function (blue surface) and true function (black wireframe), (b) optimization paths in the parameter subspace and objective values of the best solutions for the surrogate function (blue wireframe) and true function (red wireframe).

For a deeper discussion of the achieved best solutions, Figure 2a shows a comparison between the surrogate and true objective functions in the bi-dimensional subspace of the ($\rho_r/\rho_{an}$, $w_{an}/w$)-parameters, when the other parameters are fixed at the values of the solution $\boldsymbol{\mu}^*_{surr}$. More specifically, Figure 2b reports, for the particular repetition of the SLP algorithm for which the best solution $\boldsymbol{\mu}^*_{surr}$ is obtained, the optimization paths followed by the algorithm iterations in the ($\rho_r/\rho_{an}$, $w_{an}/w$)-plane (blue dots). The path convergence to the maximum of the surrogate objective function (dot on the blue wireframe surface) can be appreciated. The convergent behavior is reported also for the repetition of the GCMMA algorithm for which $\boldsymbol{\mu}^*_{true}$ is obtained. In this case, the other parameters are fixed at the values of the solution $\boldsymbol{\mu}^*_{true}$. The optimization path followed by the GCMMA algorithm iterations in the ($E_r/E_s$, $\rho_r/\rho_{an}$)-plane (red dots) can be observed to differ from that followed by the SLP algorithm. Again, the path convergence to the maximum of the true objective function (dot on the red wireframe surface) can be appreciated. Similar comparisons between the surrogate and the true objective functions, together with the different paths followed by the SLP and GCMAA algorithms are reported in Figure 3 and Figure 4 in the bi-dimensional subspace of the ($\rho_r/\rho_{an}$, $E_r/E_s$)-parameters and the ($E_r/E_s$, $w_{an}/w$)-parameters, respectively. The analyses related to the dependence on the remaining pa-rameter $v_r$ are not reported, since the two objectives can be verified to be essentially independent of the $v_r$-variations within the two-dimensional neighborhoods of the $v_r$-value of the best solutions $\boldsymbol{\mu}^*_{surr}$ and $\boldsymbol{\mu}^*_{true}$.

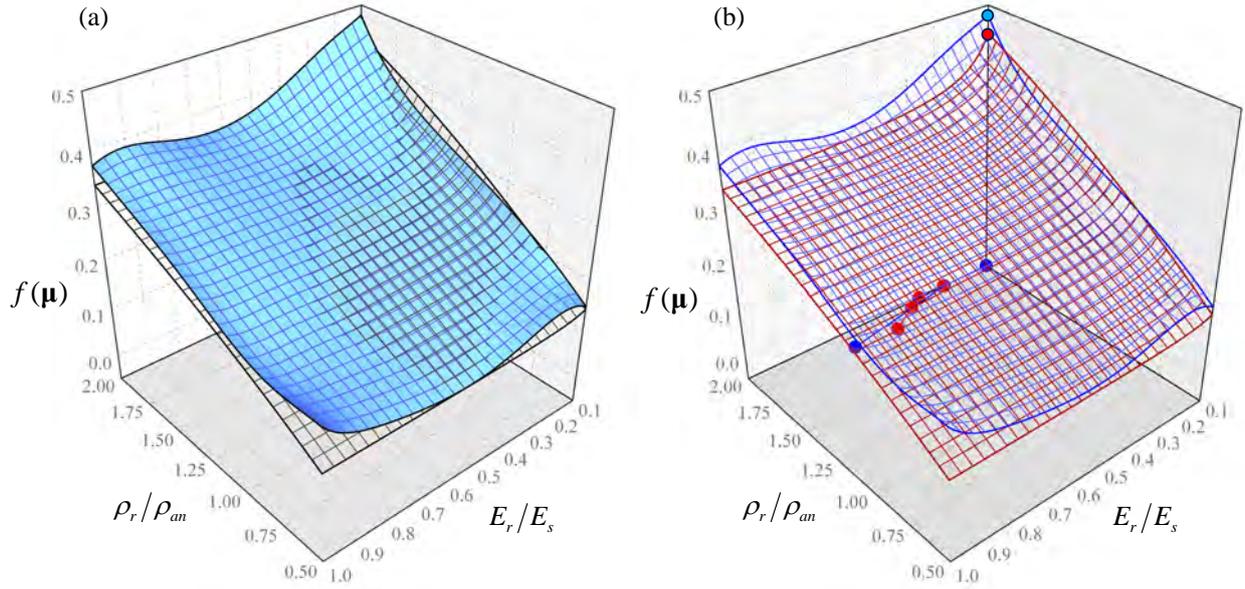

**Figure 3**. Objective functions in the bi-dimensional subspace of the admissible parameters $\rho_r/\rho_{an}$ and $E_r/E_s$: (a) surrogate function (blue surface) and true function (black wireframe), (b) optimization paths in the parameter subspace and objective values of the best solutions for the surrogate function (blue wireframe) and true function (red wireframe).

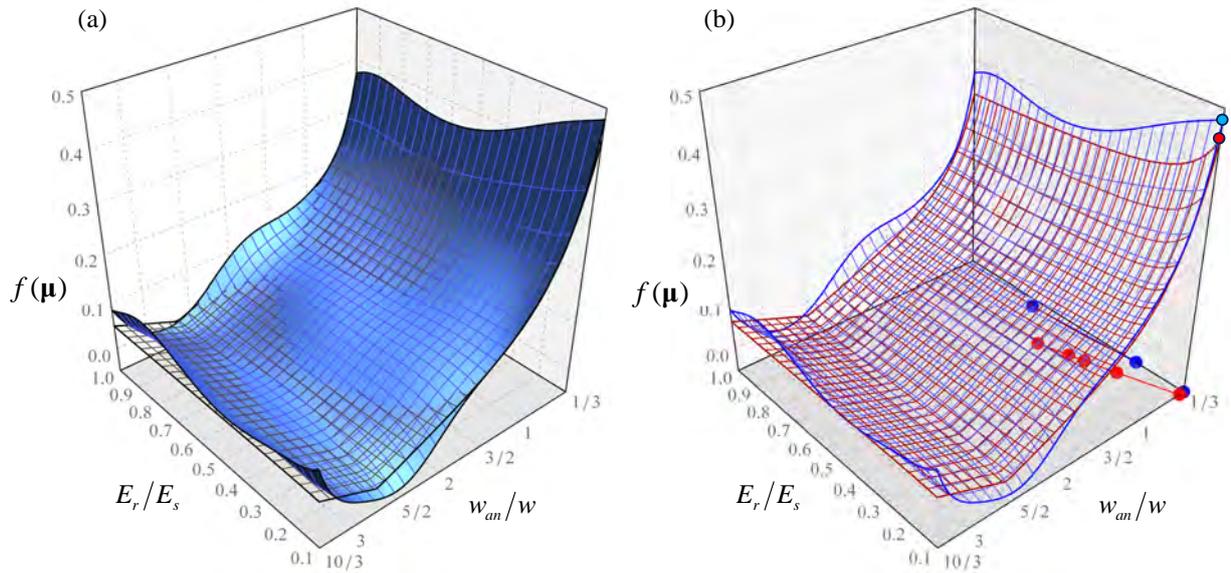

**Figure 4**. Objective functions in the bi-dimensional subspace of the admissible parameters $E_r/E_s$ and $w_{an}/w$: (a) surrogate function (blue surface) and true function (black wireframe), (b) optimization paths in the parameter subspace and objective values of the best solutions for the surrogate function (blue wireframe) and true function (red wireframe).

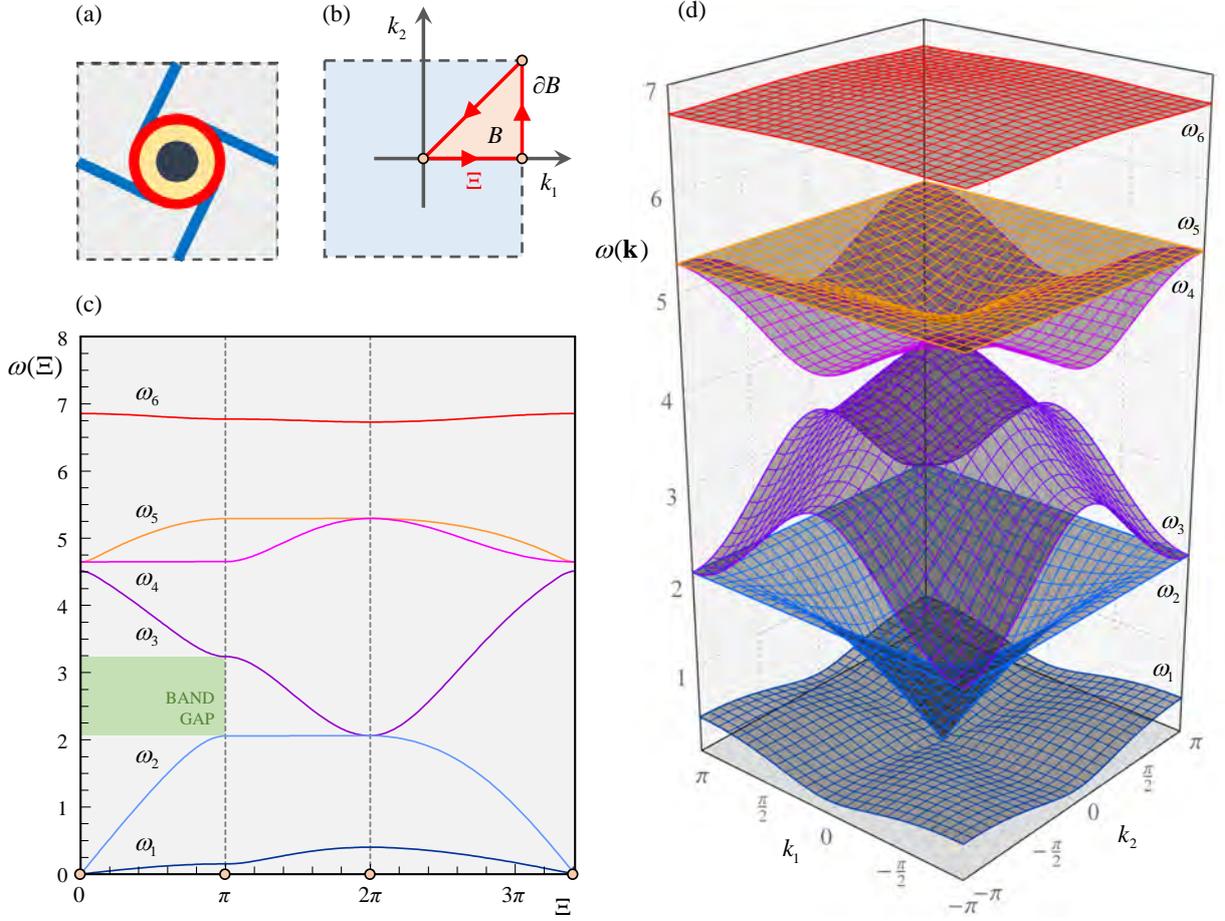

**Figure 5**. Tetrachiral metamaterial, designed according to the surrogate optimization strategy: (a), (b) periodic cell and closed boundary $\partial B$ of the subdomain $B$ for the corresponding Brillouin zone, (c) dispersion curves with the optimally-amplified band gap, (d) dispersion surfaces.

Finally, the dispersion spectrum of the tetrachiral metamaterial, designed according to the surrogate optimization strategy (best solution $\boldsymbol{\mu}^*_{surr}$ reported in Table 2), is portrayed in Figure 5. First, the boundary $\partial B$ closing the subdomain $B$ of the Brillouin zone corresponding to the periodic cell of the metamaterial is presented (Figure 5a,b). Second, the six dispersion curves composing the spectrum are illustrated, with focus on the optimally-amplified partial band gap (green region) separating the dispersion curves of the frequencies $\omega_2$ and $\omega_3$ over the horizontal segment of the boundary $\partial B$, spanned by the local abscissa $\Xi$ (Figure 5c). Lastly, the six dispersion surfaces of the designed metamaterial are reported in the entire bi-dimensional Brillouin zone spanned by the wavevector $\mathbf{k} = (k_1, k_2)$.

## 5. Conclusions and further research directions

A viable approach to apply machine-learning techniques to spectral optimization problems for periodic metamaterials has been investigated, by leveraging a surrogate optimization strategy based on a suited combination of Radial Basis Function network interpolation methods and

Quasi-Monte Carlo sequences. Motivated by the pressing operational need to significantly reduce the high computational effort involved in tackling band gap optimization problems, the surrogate optimization strategy has been verified to efficiently treat the large size of the physical matrices defining the spectral eigenproblems and the extra-fine discretization of the dispersion function domain, which traditionally require unaffordable computational resources for the evaluation of the objective functions. Specifically, numerical results have been reported for a band gap optimization problem related to the maximization of the partial band gap in the acoustic low-frequency spectrum of the periodic tetrachiral material. The superior performance of the surrogate optimization strategy (in terms of reduced number of numerical evaluations to achieve a certain optimal solution) has been clearly highlighted from the direct comparison with a more traditional approach based on the exact evaluation of the true objective function, still applicable by virtue of the relatively small number of variables to be optimized.

As major concluding remark, the machine-learning approach to spectral optimization problems is open to further promising improvements. Specifically, some feasible ideas for ongoing and future investigations can be found in the following directions:

a) a more sophisticated iterative algorithm for surrogate optimization could be applied, if an updating of the surrogate objective function is provided during the execution of the algorithm, by evaluating the true objective function on points selected by the algorithm itself;

b) Gaussian RBF networks with *variable* widths and centers could be used, since it is known that this improvement helps mitigating the curse of dimensionality for other (function approximation and optimization) problems [38],[39],[40],[41]. Similarly, feedforward neural networks with *sigmoidal* computational units could be employed, since they offer similar guarantees [38],[42]. Together with other classes of approximators, both these kinds of networks are applied extensively by the so-called Extended Ritz Method [43],[44],[45], allowing to get suboptimal solutions to functional optimization problems with guarantees on their performance. Likewise Gaussian RBF networks, also feedforward neural networks with sigmoidal computational units can be applied to perform surrogate optimization [46];

c) fixed the total computational time, the acceptable precision in the evaluation of the true objective function on the training set could be decreased. This could be done, for instance, by using a coarser grid for the closed boundary of the subdomain of the Brillouin zone, or by decreasing the precision in the eigenvalue computations. By doing this, the generation of each (less precise) training example would require less time, hence a larger number of training examples could be exploited to construct the surrogate objective function;

d) still in order to use a coarser grid and allow at the same time a precise evaluation of the band gap amplitude, machine-learning methods could be applied for dispersion curve identification in the presence of curve intersections [47].

As final perspective, the development and the application of efficient surrogate optimization algorithms to band gap optimization problems can be argued to be an important preliminary step for the optimal design of more multi-field electro-mechanical models representing *smart* (e.g., automatically tunable) *filters*, characterized by a much larger number of (tunable) parameters. For such problems, an online re-optimization of the parameters as a consequence of a change in the operating conditions could be extremely time consuming if it were based on exact evaluations of the true band gap objective functions (especially if a large number of such evaluations were needed). Then, surrogate optimization may represent a valid alternative to exact computations, possibly providing good suboptimal solutions in reasonable amounts of time.

**Conflict of Interest**

The authors declare that they have no conflict of interest.

**Acknowledgments**

The authors acknowledge financial support of the (MURST) Italian Department for University and Scientific and Technological Research in the framework of the research MIUR Prin15 project 2015LYYXA8, "Multi-scale mechanical models for the design and optimization of micro-structured smart materials and metamaterials". The authors also acknowledge financial support by National Group of Mathematical Physics (GNFM-INdAM).

**Appendix**

Some specific mathematical concepts and the related technical notations employed in the paper are reported in further detail. In particular, Appendix A describes mesh-free interpolation methods, Appendix B provides details on surrogate optimization, and Appendix C is concerned with iterative optimization algorithms.

**Appendix A: Mesh-free interpolation methods**

A *Radial Basis Function* (*RBF*) network interpolation method [28] based on $N_c$ computational units is characterized by an interpolant $\hat{f}: \Omega \subset \mathbb{R}^p \to \mathbb{R}$ of the form

$$\hat{f}(\mathbf{x}) = \sum_{i=1}^{N_c} c_i \varphi_i (\mathbf{x} - \mathbf{x}_i), \qquad (3)$$

where, for $i=1,\ldots,N_c$, the $c_i \in \mathbb{R}$ are the coefficients of the linear combination, the $\varphi_i: \mathbb{R} \to \mathbb{R}$ are radial basis functions (i.e., functions whose values depend only on the distances of their arguments from the origin), and the $\mathbf{x}_i \in \mathbb{R}^p$ are called *centers*. RBF network interpolation methods are examples of *mesh-free methods* that, differently from *mesh-based ones*, do not require the presence of connections between their nodes, but are rather based on the interaction between each node and all its neighbors. As a particular case, the *Gaussian RBF network interpolation method* exploits an interpolant of the form

$$\hat{f}(\mathbf{x}) = \sum_{i=1}^{N_c} c_i \exp\left(-\frac{\|\mathbf{x}-\mathbf{x}_i\|_2^2}{2\sigma_i^2}\right), \quad (4)$$

where $\|\mathbf{x}-\mathbf{x}_i\|_2$ denotes the Euclidean norm of the vector $\mathbf{x}-\mathbf{x}_i$, and the coefficients $\sigma_i > 0$ are called widths. In the paper, a Gaussian RBF network interpolation method with fixed centers and the same widths $\sigma_i = \sigma$ is used, where the coefficients $c_i$ are chosen in such a way to make, for a function $f: \Omega \subset \mathbb{R}^p \to \mathbb{R}$ to be interpolated, the error associated with its interpolation by $\hat{f}$ equal to zero on a finite set of input training points in $\mathbb{R}^p$. In the following, such a set is taken to be coincident with the set of $N_c$ centers. Consequently, the coefficients $c_i$ are obtained by solving the linear system

$$\mathbf{Ac} = \mathbf{f}, \quad (5)$$

where $\mathbf{A} \in \mathbb{R}^{N_c \times N_c}$ is called *interpolation matrix*, whose generic element, in the case of Gaussian RBF networks, has the form $A_{ij} = \exp\left(-\|\mathbf{x}_i - \mathbf{x}_j\|_2^2 / (2\sigma^2)\right)$, whereas $\mathbf{f} \in \mathbb{R}^{N_c}$ is a column vector collecting the values assumed by the function $f$ on the input training points, and $\mathbf{c} \in \mathbb{R}^{N_c}$ is the column vector collecting the unknown coefficients of the linear combination.

Typically, radial basis functions used in RBF network interpolation methods are chosen to be strictly positive-definite. It is recalled here that a symmetric function $g: \mathbb{R} \to \mathbb{R}$ is called *strictly positive-definite* if, for every positive integer $n$ and every collection of $n$ distinct elements $x_1, ..., x_n \in \mathbb{R}$, the symmetric matrix $\mathbf{G} \in \mathbb{R}^{n \times n}$ having elements of the form $\mathbf{G}_{ij} = g(x_i - x_j)$ is positive definite. Since the interpolation matrix $\mathbf{A}$ belongs to this matrix class, strict positive-definiteness of a fixed (i.e., independent from the index $i$) radial basis function guarantees the nonsingularity of the associated interpolation matrix, hence existence and uniqueness of the associated interpolant, for every choice of the training set, provided the input training points are distinct (see also [28], Chapter 3). In this way, indeed, the solution of equation (5) is

$$\mathbf{c} = \mathbf{A}^{-1}\mathbf{f}. \quad (6)$$

When the training points do not coincide with the centers, the equation (5) may not be solvable and an approximate (typically not interpolating) solution must be searched. For instance the solution $\mathbf{c}^+ = \mathbf{A}^+ \mathbf{f}$ (where $\mathbf{A}^+$ is the Moonre-Penrose pseudo-inverse of $\mathbf{A}$) or the Tikhonov-regularized solution $\mathbf{c}^\lambda = \left(\mathbf{A}^T \mathbf{A} + \lambda \mathbf{I}\right)^{-1} \mathbf{A}^T \mathbf{f}$, where $\lambda > 0$ is a suitable regularization parameter, can be achieved. However, for the kind of band gap optimization problems considered in this paper, interpolation is preferred, since there is no noise in the training set.

Despite the uniqueness of the interpolation for every choice of the width parameter $\sigma$, different choices provide different interpolants, with varying quality of the approximation outside the training set. A possible way to select a suitable value for $\sigma$, aimed to prevent overfitting, is *leave-one-out cross-validation* [28], which in the context of RBF network interpolation is also called *Rippa method*. According to this method, for each choice of $\sigma$, $N_c$ different RBF network interpolants are constructed starting from $N_c$ different – but highly overlapping – training sets of size $N_c - 1$, where each of them is obtained by removing a different training point from the full training set of size $N_c$. Then, for the $k$-th such interpolant ($k = 1, ..., N_c$), the approximation error $\varepsilon_k$ is evaluated on the unique point that has been excluded from its training set. All these errors are collected in a column vector $\boldsymbol{\varepsilon} \in \mathbb{R}^{N_c}$, whose maximum norm is minimized with respect to $\sigma$ in order to choose the width optimally. Interestingly, the evaluation of each $\varepsilon_k$ does not require the actual training of the $k$-th interpolant, since it is possible to prove [29] that such error has the expression

$$\varepsilon_k = \frac{c_k}{\left(A^{-1}\right)_{k,k}}, \tag{7}$$

where $c_k$ is the $k$-th element of the column vector $\mathbf{c}$ of coefficients of the linear combination defining the interpolant trained on the whole training set, and $\mathbf{A}$ is the associated interpolation matrix (still dependent on the whole training set). Finally, the output of the leave-one-out cross-validation is made of:

    a) the optimal choice of $\sigma$ and
    b) the RBF network interpolant trained on the whole training set, for that value of $\sigma$.

If $N_c$ is large, the interpolant is expected to be very similar those trained – for the same choice of $\sigma$ – on the $N_c$ smaller training sets of size $N_c - 1$, due to their high overlap.

**Appendix B: Surrogate optimization**

*Surrogate optimization* replaces the objective function of an optimization problem with a suitable interpolant (or more generally, with a suitable approximating function), which is called surrogate objective function [20]. Once a suitable optimization algorithm is chosen to solve the resulting surrogate optimization problem, the performance of surrogate optimization can be measured in terms of:

a) the quality of the suboptimal solution it provides, expressed in terms of the value assumed by the surrogate objective function at the suboptimal solution, and by its comparison with the value assumed by the true objective function;
b) the total computational time needed by the surrogate optimization algorithm to find the suboptimal solution;
c) its number of evaluations of the true objective function.

In the paper, a fixed surrogate objective function is considered, i.e., the interpolant of the true objective function is constructed in the initialization phase of the surrogate optimization algorithm and remains unchanged. More advanced surrogate optimization algorithms update their interpolants during their iterations, by exploiting additional information coming from the evaluation of the true objective function on points selected adaptively by the algorithm itself [21].

Constructing an interpolant requires the choice of a suitable input training set, on which the function to be interpolated is evaluated. In this way, suitable coefficients of the linear combination defining the interpolant are obtained (possibly together with additional coefficients, like the width parameter in Gaussian RBF network interpolation). Two possible choices for the training set are provided by:

a) *Quasi-Monte Carlo* discretization, according to which the input training set is originated as a subsequence of a Quasi-Monte Carlo (deterministic) sequence, and

b) *Monte Carlo* discretization, which exploits, instead, a realization of a Monte Carlo (random) sequence.

Both kinds of sequences are used extensively in numerical integration [29]. In that context, Quasi-Monte Carlo sequences are often preferred to realizations of Monte Carlo ones, because the former are typically able to cover the domain $\Omega$ in a more uniform way than the latter (see, e.g., [17] for an illustrative comparison). In other words, points generated from Quasi-Monte Carlo sequences tend to form clusters. This issue depends on the Monte Carlo algorithm that, once a specific point belonging to a Monte Carlo sequence has been generated, keeps no memory of it to generate the successive point of the sequence (being such points realizations of independent vector-valued random variables). An additional feature of Quasi-Monte Carlo sequences is that, since they are deterministic, they generate perfectly reproducible point sets. However, this can be achieved also by Monte Carlo discretization, if pseudo-random (deterministic) sequences, whose statistical properties are similar to the ones of Monte Carlo sequences, are exploited.

The use of Quasi-Monte Carlo sequences (and of the point sets associated with their subsequences) also in function interpolation can be justified using the concept of dispersion. Let $X = \{\mathbf{x}_1,...,\mathbf{x}_{N_c}\} \subset \Omega$ be a finite subset of the domain. Then, its *dispersion* (or *fill distance*) is defined as

$$h_{X,\Omega} = \sup_{\mathbf{x} \in \Omega} \min_{\mathbf{x}_j \in X} \left\| \mathbf{x} - \mathbf{x}_j \right\|_2. \tag{8}$$

According to [28] (Theorem 14.5), a large class of RBF network interpolants $\widehat{f}_X : \mathbb{R}^p \to \mathbb{R}$ with centers and input training points coincident with the elements of $X$ is characterized by the following upper bound on the error (in the sup norm) in the approximation of a function $f : \mathbb{R}^p \to \mathbb{R}$ satisfying a suitable smoothness condition

$$\sup_{\mathbf{x}\in\Omega}\left|f(\mathbf{x})-\widehat{f}_X(\mathbf{x})\right|\leq Ch_{X,\Omega}^k\|f\|_H,\qquad(9)$$

where $C$ is a positive constant (which does not depend on $f$ and on the choice of the fixed radial basis function), $k$ is a lower bound on the degree of smoothness of the computational units, and $\|f\|_H$ is the norm of $f$ in a suitable function space $H$, which is associated with the specific choice of the fixed radial basis function. Of course, to reduce the value of the upper bound above on the approximation error, one possibility is to choose a low-dispersion sequence.

It has to be observed that the bound (9) requires a smoothness assumption on the function $f$, namely, its belonging to the space $H$. When this does not hold, the domain $\Omega$ can be replaced with a subdomain on which $f$ is smooth, or $f$ can be replaced with a smooth approximation (possibly coincident with $f$ on $X$). The bound (9) is then applied to the $f$-approximation. In this case, the presence of the error associated with the additional approximation step must be taken into account.

Interestingly, when the domain $\Omega$ is the $p$-dimensional hypercube $[0,1]^p$, the dispersion $h_{X,[0,1]^p}$ can be bounded from above in terms of another property of the set $X$ (which can be also related to the associated sequence $\mathbf{x}_1,...,\mathbf{x}_{N_c}$), called *discrepancy* and defined as

$$D_X=\sup_{G\in\Im}\left|\frac{S(G)}{N_c}-\prod_{i=1}^{N_c}(b_i-a_i)\right|,\qquad(10)$$

where $\Im$ is the family of all subsets of $[0,1]^p$ having the form $G=\prod_{i=1}^{N_c}[a_i,b_i)$, and $S(G)$ is the cardinality of the intersection $X\cap G$. Then, the upper bound on the dispersion (see, for instance Theorem 6.6 in [29]) is

$$h_{X,[0,1]^p}\leq p^{\frac{1}{2}}\left(D_X\right)^{\frac{1}{p}},\qquad(11)$$

hence it zeroes when the discrepancy vanishes. Since Quasi-Monte Carlo sequences are characterized by low discrepancy, also their dispersions are low. This justifies the use of Quasi-Monte Carlo sequences (e.g., the *Sobol' sequence*) also in function interpolation.

To conclude, it is worth remarking that upper bounds on the approximation error of the function $f$ through its interpolant $\widehat{f}_X$, similar to that expressed in the sup norm and defined in equation (9), allow also to bound from above the error in the approximation of the maximum value of $f$ in terms of the maximum value of $\widehat{f}_X$. Indeed, assuming that the respective maxima exist, equation (9) implies

$$\left|\max_{\mathbf{x}\in\Omega}f(\mathbf{x})-\max_{\mathbf{x}\in\Omega}\widehat{f}_X(\mathbf{x})\right|\leq Ch_{X,\Omega}^k\|f\|_H.\qquad(12)$$

Similar guarantees cannot be obtained if one starts, instead, from upper bounds on the approximation error expressed in Lebesgue-space norms like the $L_2$ norm. Analogous remarks hold for other optimization problems to which Gaussian RBF networks have been applied [38],[40].

**Appendix C: Iterative optimization algorithms**

Both the original and surrogate band gap optimization problems are characterized by the presence of continuous variables to be optimized, a nonlinear objective function, and linear and/or nonlinear constraints. To solve them approximately, suitable iterative optimization algorithms can be applied. In general, the number of iterations being the same, the performance of each such algorithm is expected to produce better suboptimal solutions if it is applied to the original optimization problem rather than to the surrogate one. This is motivated by the absence of the additional approximation error associated with the use of the surrogate objective function. Nevertheless, the iterations of the algorithm are faster when it is applied to a surrogate optimization problem with a fixed surrogate objective function, since such function – once it has been constructed – is cheaper to evaluate. This advantage holds also for its partial derivatives, if they are used during the iterations of the algorithm. For instance, partial derivatives of a Gaussian RBF network interpolant can be computed in closed form, once the values of the parameters of the interpolant have been chosen. So, even better suboptimal solutions may be produced starting from such surrogate objective function, the total computational time being the same. This possibility is particularly important when there are upper bounds on the available total computational time. It is worth mentioning that these arguments have been proved in [48] for the related problem of regression, whose investigation can be considered as a preliminary step for the analysis of surrogate optimization (indeed, it involves only the construction of the approximating function, but not its successive optimization). Possible choices for the iterative optimization algorithm, examined in this paper are:

a) *Method of Moving Asymptotes* (*MMA*) [33] and its *Globally Convergent* upgrade (*GCMMA*) [34]. The *GCMMA* algorithm – which was developed originally with the specific aim of solving optimization problems arising in structural engineering – replaces the initial (either the original, or the surrogate) optimization problem with a sequence of approximating optimization subproblems, which are easier to solve, though being still nonlinearly constrained. In each subproblem, the objective and constraining functions of the initial optimization problem are replaced by suitable approximating functions. From a certain viewpoint, this can be also considered as a form of surrogate optimization. However, such approximations are characterized by a much simpler functional form than Gaussian RBF network interpolation. In any case, the resulting subproblems are easier to solve than the initial optimization problem, e.g., because GCMMA applies *separable* approximators, which are summations of functions depending on a single real argument. Nevertheless, just because of their simplicity, the quality of such approximations may be worse than those of Gaussian RBF network interpolation, if a sufficiently large number of basis functions is used. Moreover, Gaussian RBF network interpolation guarantees a zero approximation error on the training set. Still, GCMMA is globally convergent in the sense that, for each possible initialization of the set of variables to be optimized, the algorithm was proved in [34] to converge to a stationary point of the initial optimization problem (assuming twice-continuous differentiability of the objective and constraining functions; in practice, convergence is often observed also for less smooth functions). Finally, it can be recalled that the expression *moving asymptotes* reflects that, in each iteration, GCMMA exploits approximating functions characterized by suitable asymptotes, which typically change (move) from one iteration to the successive one;

b) *Sequential Linear Programming* (*SLP*) [19]. Also in this case, the initial optimization problem is replaced by a sequence of simpler approximating subproblems. However, differently from GCMMA, at each iteration of SLP, linearizations of the objective and constraining functions at the current suboptimal solution are exploited. Moreover, additional box constraints are inserted, which define a so-called *trust region* inside which the approximation error due to the linearizations above is guaranteed to be small (such a guarantee can be obtained, e.g., by comparing the actual derivatives of the objective and constraining functions on selected points belonging to the trust region with their constant approximations coming from the linearizations above). Each resulting optimization subproblem is even easier to solve than those occurring in the application of GCMMA. Indeed, the simplex algorithm can be applied, or – if the number of variables to be optimized is not too large – the objective function can be evaluated only on the vertices of the admissible region of each subproblem, which are simplexes. Nonetheless, convergence may be slower in practice with respect to GCMMA, because of the extreme simplicity of the linear approximations, and of the need to choose a sufficiently small trust region, in order to guarantee a small approximation error in the linearizations. However, a possible way to improve the rate of convergence is obtained by exploiting an *adaptive trust region* (i.e., a trust region having an adaptive size), which can be constructed by enlarging (decreasing) the size of the trust region if the current linear approximations are *good* (*bad*). Still, a lower and upper bound on the size of the adaptive trust region are needed, in order to avoid such size growing or shrinking indefinitely, possibly generating numerical errors.

Iterative optimization algorithms need suitable *termination criteria* [19]. The simplest criterion consists in stopping the algorithm after a given number $N_i$ of iterations. However, if different algorithms characterized by different computational times per iteration are used, then a fair approach to compare their performances consists in stopping each algorithm after the same total computational time. Other termination criteria involve, e.g., the comparison of the values assumed by the (true or surrogate) objective function in correspondence of the suboptimal solutions generated in consecutive iterations of the optimization algorithm, or the comparison between the values assumed by its gradient and the all-zero vector (limiting for simplicity the discussion to the case of unconstrained optimization).

Finally, it is important to recall that iterative optimization algorithms for nonlinearly constrained optimization problems with nonlinear objective function and continuous variables to be optimized typically find only *local maxima* (the same GCMMA, though globally convergent, is guaranteed to converge only to a *local stationarity point*, which could even not coincide with a local maximum point). In order to increase the probability of finding the global maximum or a good local maximum (i.e., one whose value of the objective function is near the global maximum), several repetitions of each iterative optimization algorithm can be performed, starting from its different initializations, possibly generated using a Quasi-Monte Carlo subsequence (this is called a Quasi-Monte Carlo multi-start initialization approach). In this way, indeed, the probability of starting the iterative optimization algorithm increases – in at least one repetition – either from the *basin of attraction* of a global maximum, or from that associated with a sufficiently good local maximum [43].